\documentclass[12pt]{iopart}

\usepackage{graphicx}
\begin{document}

\title[S. Longhi, Gap solitons in metamaterials ]{Gap solitons in metamaterials}

\author{S. Longhi\dag}

\address{\dag\ Dipartimento di Fisica and Istituto di Fotonica e Nanotecnologie del CNR, Politecnico di Milano, Piazza L. da
Vinci 32,  I-20133 Milan, Italy }

\begin{abstract}
 Electromagnetic localization and existence of gap
solitons in nonlinear metamaterials, which exhibit a stop band in
their linear spectral response, is theoretically investigated. For
a self-focusing Kerr nonlinearity, the equation for the
 electric field envelope with carrier frequency in the stop band - where the
magnetic permeability $\mu(\omega)$ is positive and the dielectric permittivity
$\epsilon(\omega)$ is negative -  is described by a nonlinear Klein-Gordon equation
with a dispersive nonlinear term. A family of standing and moving localized waves for
both electric and magnetic fields is found, and the role played by the nonlinear
dispersive term on solitary wave stability is discussed.
\end{abstract}






\maketitle

\newpage

\section{Introduction}
The study of electromagnetic propagation in metamaterials, i.e.
artificially constructed media showing negative effective
dielectric permittivity $\epsilon(\omega)$ and magnetic
permeability $\mu(\omega)$, has received a tremendous and
increasing interest in the past few years especially after the
fabrication of microstructured materials showing negative
refraction at microwaves \cite{Smith00,Shelby01,Pendry99}.
Electromagnetic wave propagation in left-handed (LF) materials in
the linear regime has been extensively investigated since the
pioneering work by Veselago \cite{Veselago67} (see, e.g.,
\cite{varilineare1,varilineare2,varilineare3,varilineare4} and
references therein), and many unusual effects with a potential
impact in different fields of science have been predicted, the
most notably one being the realization of a "perfect" lens
\cite{Pendry00} which has raised a debate among the scientific
community \cite{vari}. Only very few papers
\cite{Agranovich04,Shadrivov04} have been so far devoted to study
{\em nonlinear} electromagnetic propagation in metamaterials
despite they are expected to exhibit nontrivial nonlinear
properties \cite{Zharov03}. In particular, surface-polariton
solitons have been studied in metamaterial interfaces assuming a
simple Kerr-type nonlinearity for the dielectric response of the
medium \cite{Shadrivov04}. An interesting property of composite
metamaterials, which is usually found in {\it periodic} media, is
the existence of a frequency stop band in their linear spectral
response \cite{Smith00}, corresponding to negative values of
either $\epsilon(\omega)$ or $\mu(\omega)$. As compared to most
common band gaps found in periodic media, where forbidden
frequencies arise due to multiple Bragg scattering, in composite
lossless LH materials forbidden frequencies exist because of the
special dispersion curve for the effective permittivity and
permeability, which have {\it opposite} sign in a spectral
interval which separates allowed (propagative) regions of LH
(i.e., $\epsilon, \mu <0$) and right-handed (RH, i.e. $\epsilon,
\mu >0$)
 waves. Experimental evidence for such band gaps  at microwaves was reported
 in Ref.\cite{Smith00}
 using a microstructured medium made of a periodic array of conducting split ring resonators
and wires \cite{note0}. A rather universal feature of {\it
nonlinear periodic} media exhibiting a frequency stop band in
their linear dispersive response is the existence of
self-transparent gap soliton envelopes supported by the
nonlinearity \cite{deSterke94}, which have been extensively
investigated in the fields of nonlinear optics
\cite{deSterke94,Christodoulides89_a,Christodoulides89_b,Christodoulides89_c}
and matter waves \cite{Zobay99_a,Zobay99_b} and usually modelled
by a generalized massive Thirring model. The existence of a
frequency stop band in metamaterials connecting RH and LH
propagative regions raises the question whether, in presence of
nonlinear effects, gap solitons can be supported in
microstructured media, despite Bragg scattering effects do not
play any role and the medium behaves as a homogeneous one
\cite{note0}. In this Letter it is shown indeed that envelope gap
solitons can be supported in lossless metamaterials exhibiting a
cubic dielectric nonlinearity when the carrier frequency of the
fields is tuned inside the stop band separating LH and RH
propagative spectral regions. In Sec. 2 a nonlinear envelope
equation for the electric field is derived starting from Maxwell's
equations in the limit of a narrow spectral band gap. The
resulting equation is a complex nonlinear Klein-Gordon equation
(NLKGE) with a {\it dispersive} cubic term, which arises due the
strong dispersion of the magnetic permeability for linear waves.
Solitary wave solutions to this equation are studied analytically
and numerically in Sec.3. In particular, a two-parameter family of
steady or moving bright solitary waves for both electric and
magnetic fields is found. Numerical simulations indicate also that
these solitary waves are stable.

\section{Electromagnetic wave propagation in metamaterials with a cubic nonlinearity:
derivation of the nonlinear envelope equation}
 Let us consider propagation of electromagnetic plane waves in a
dispersive medium, with frequency-dependent relative dielectric
permittivity $\epsilon(\omega)$ and magnetic permeability
$\mu(\omega)$, which exhibits an instantaneous cubic (Kerr-type)
nonlinearity in its dielectric  response. Assuming
quasi-monochromatic electric and magnetic fields with carrier
frequency $\omega_0$ polarized along the transverse $x$ and $y$
directions,\\
$\mathbf{E}(z,t)=\mathcal{E}(z,t) \exp(i \omega_0t) \mathbf{u}_x
+c.c.$ and\\
$\mathbf{H}(z,t)=\mathcal{H}(z,t) \exp(i \omega_0t)
\mathbf{u}_y +c.c.$, a nonlinear polarization term of the medium
$\mathbf{P}_{NL}=\epsilon_0 \chi^{(3)} (\mathbf{E} \cdot
\mathbf{E}) \mathbf{E}$ ($\chi^{(3)}>0$ for a self-focusing
nonlinearity), from Maxwell's equation the following coupled-mode
equations for the slowly-varying electric and magnetic field
envelopes $\mathcal{E}$ and $\mathcal{H}$ can be derived in the
rotating-wave approximation:
\begin{eqnarray}
\partial_{z} \mathcal{E} & = & -i \mu_0 (\omega_0-i \partial_t) \mu (\omega_0 -i \partial_t)
\mathcal{H}\\
\partial_{z} \mathcal{H} & = & -i \epsilon_0 (\omega_0-i \partial_t) \epsilon (\omega_0 -i \partial_t) \mathcal{E}
 -3i \omega_0 \epsilon_0 \chi^{(3)} |\mathcal{E}|^2 \mathcal{E},
\end{eqnarray}
where $\epsilon_0$ and $\mu_0$ are the vacuum dielectric permittivity and magnetic
permeability, and the operators $\epsilon(\omega_0-i \partial_t)$, $\mu(\omega_0-i
\partial_t)$ are defined, as usual, by the power expansion of $\epsilon(\omega_0+\Omega)$ and
$\mu(\omega_0+\Omega)$ at around $\Omega=0$ with the substitution $\Omega \rightarrow
-i \partial_t$ (see, e.g., \cite{Newell}). A single equation for the electric field
envelope $\mathcal{E}$ can be derived from Eqs.(1) and (2) and reads explicitly:
\begin{equation}
\partial^{2}_{z} \mathcal{E}+k^2(\omega_0-i \partial_t) \mathcal{E}= -3 \omega_0
\epsilon_0 \mu_0 \chi^{(3)} (\omega_0 -i \partial_t) \mu(\omega_0-i \partial_t)
|\mathcal{E}|^2 \mathcal{E},
\end{equation}
where we have set $k^2(\omega) \equiv \epsilon_0 \mu_0 \omega^2 \epsilon(\omega)
\mu(\omega)$. For a composite metamaterial made by an array of conducting nonmagnetic
split-ring resonators and continuous wires, we may assume the following general form
for the permeability and permittivity (see, e.g. \cite{Smith00} and references
therein):
\begin{equation}
\epsilon(\omega)=1- \frac{\omega_{p}^{2}}{\omega(\omega-i \gamma_{\epsilon})}  \;\;
,\;\; \mu(\omega)=1-\frac{F \omega^2}{\omega^2-\omega_{m}^2-i \gamma_{\mu} \omega},
\end{equation}
where $\omega_p$, $\omega_m$, $F$, $\gamma_{\epsilon}$ and
$\gamma_{\mu}$ can be tuned by changing the geometrical and
physical parameters of the microstructures forming the composite
medium \cite{Smith00,Pendry99}. For an ideal lossless medium, one
has $\gamma_{\epsilon}=\gamma_{\mu}=0$. Referring to this ideal
case and assuming a plasma frequency $\omega_p$ larger than
$\overline \omega = \omega_m/(1-F)^{1/2}$, linear waves of Eq.(3)
are evanescent ($k^2(\omega)<0$) in the spectral interval
$\overline \omega< \omega < \omega_p$ (band gap) , where
$\mu(\omega)>0$ but $\epsilon(\omega)<0$, whereas they are
propagative ($k^2(\omega)>0$) for $\omega>\omega_p$ and
$\omega<\overline \omega$ (down to the resonance $\omega_m$), with
a LH [RH] behavior of the material on the left [right] side of the
band gap. A typical behavior of $\epsilon (\omega)$ and $\mu
(\omega)$, for parameter values compatible with recent
experimentally fabricated structures \cite{Smith00}, is shown in
Fig.1. Let us now assume a reference carrier frequency $\omega_0$
for the fields at the center of the band gap, i.e. $
\omega_0=(\omega_p+ \overline \omega)/2$, and assume a
sufficiently narrow band gap ($\omega_p-\overline \omega \ll
\omega_0$) such that the dispersion curves $\epsilon(\omega)$ and
$\mu(\omega)$ can be expanded up to leading order in
$\omega-\omega_0$ at around the bandgap region, i.e.
$\epsilon(\omega) \simeq \epsilon^{'}_{0}(\omega-\omega_0-\Delta)$
and $\mu(\omega) \simeq \mu^{'}_{0}(\omega-\omega_0+\Delta)$,
where $\Delta=(\omega_p-\overline \omega)/2$ is the half width of
the stop band and $\epsilon^{'}_{0} \equiv (\partial \epsilon /
\partial \omega)_{\omega_0}$, $\mu^{'}_{0} \equiv (\partial \mu /
\partial \omega)_{\omega_0}$ are real-valued for the lossless
medium and always positive for causality \cite{Landau}. Such approximation corresponds
to assume a parabolic behavior of $k^2 (\omega)$ versus $\omega$, which is reasonable
as shown in Fig.1(b). With these assumptions, for slowly-varying envelopes ( $(1/
\omega_0) \partial_t \sim (\Delta / \omega_0) \ll 1 $) and introducing the
dimensionless variables $t'=\Delta t$, $z'=\Delta z/v_g$, $\psi=[3
\chi^{(3)}/(\epsilon_{0}^{'} \Delta) ]^{1/2} \mathcal{E}$, and $ \varphi= [3 \chi^{(3)}
\mu_0 /( \epsilon_0 \epsilon_{0}^{'} \Delta) ]^{1/2} \mathcal{H}$, where $v_g \equiv
(\epsilon_0 \mu_0 \omega_{0}^{2} \epsilon_{0}^{'} \mu_{0}^{'})^{-1/2}$, Eqs.(1) and (2)
read:
\begin{eqnarray}
\partial_{z'} \psi & = & i {\sqrt  \frac{\mu_{0}'}{\epsilon_{0}^{'}}} \left( -1+i \partial_{t'}
\right) \varphi \\
\partial_{z'} \varphi & = &  i {\sqrt \frac{\epsilon_{0}^{'}}{\mu_{0}^{'}}} \left( 1+i \partial_{t'}
\right) \psi -i {\sqrt \frac{\epsilon_{0}^{'}}{\mu_{0}^{'}}}
|\psi|^2 \psi .
\end{eqnarray}
Finally, from Eqs.(5) and (6) the following nonlinear equation for
the normalized electric field envelope $\psi$ can be obtained:
\begin{equation}
\left( \partial^{2}_{z'} - \partial^{2}_{t'} -1 + |\psi|^2 \right)
\psi=i
\partial_{t'} \left( |\psi|^2 \psi \right)
\end{equation}
which is the basic equation describing nonlinear wave propagation
in Kerr-type nonlinear metamaterials at frequencies close to (or
inside) the stop band separating the LH and RH propagative
spectral regions of the medium.

\section{Gap solitons}
In this section we focus our analysis on the existence and
stability of stationary and moving solitary waves to the nonlinear
envelope equation derived in the previous section, Eq.(7). From a
mathematical viewpoint, Eq.(7) is a complex NLKGE which differs
from the most usual ones encountered in other physical fields
(see, e.g. \cite{Laedke82_a,Laedke82_b,Hawrylak84_a,Hawrylak84_b})
 due to the presence of a {\it nonlinear dispersive term}, represented by the right hand side
term in Eq.(7). The physical reason for the appearance of a
nonlinear dispersive term in the envelope equation despite the
instantaneous (i.e. non-dispersive) nature of the Kerr
nonlinearity is due to the fact that close to the bandgap ($\omega
\sim \omega_0$) the magnetic permeability $\mu(\omega)$ is small
and its frequency dependence can not be neglected, leading to the
appearance of the nonlinear dispersive term in Eq.(7)
\cite{note2}. Using a common terminology adopted for solitary
waves in nonlinear periodic media, we will call stationary or
moving localized waves to Eq.(7) gap solitions. However, the
reader should be aware that gap solitons in metamaterials, as
described by Eq.(7), have a different physical origin and should
not be confused with usual Bragg solitons in nonlinear periodic
media, which are described by a generalized massive Thirring model
for counterpropagating fields. In fact, as we already pointed out
\cite{note0}, despite artificial metamaterials show a microscopic
periodic structure, they behave as homogeneous media and Bragg
scattering does not occur. In this sense, gap solitons to Eq.(7)
should be regarded as a kind of self-transparency solitary waves
rather than true Bragg solitons, thought no coherent effects of
the medium are involved.\\
In order to find an analytical form of gap solitons, let us first
observe that, if the dispersive nonlinear term were negligible,
Eq.(7) would reduce to the well-known $\psi^4$ NLKGE; existence
and stability of bright solitary waves for such an equation were
previously studied in Ref.\cite{Laedke82_a,Laedke82_b}. Steady
solitary waves read explicitly\\
$\psi(z',t')=[2(1-\Omega^2)]^{1/2} {\rm
sech}[(1-\Omega^2)^{1/2}z'] \; \exp(i \Omega t')$, where $\Omega$
is a free-family parameter ($|\Omega|<1$ for existence), which
measures the frequency offset of the soliton from the band gap
center. By exploiting the Lorentz invariance of the $\psi^4$
NLKGE, a family of moving gap solitons, with an arbitrary velocity
$v$ satisfying the condition $|v|<1$ ({\it slow} gap solitons),
can be then generated. A linear stability analysis of such
solitary waves shows that they are linearly stable provided that
$1/ \sqrt 2<|\Omega|<1$ \cite{Laedke82_b}. If -as it is our case -
the nonlinear dispersive term in Eq.(7) can not be neglected, an
explicit analytical form of solitary waves can not be obtained in
general, and Lorentz invariance is broken. We can nevertheless
look for moving bright solitary waves of Eq.(7) in the form
\cite{note1} $\psi(z',t')=F(z'-v t') \exp[i(\Omega t'-Qz')]$ with
$F(\xi) \rightarrow 0$ as $ \xi \rightarrow \infty$, where $v$ is
the solitary wave speed and $\Omega$, $Q$ are real-valued
parameters; with the further constraint $Q=v \Omega$, the envelope
$F$ can be found as a homoclinic loop emanating from the unstable
solution $F=0$ of the equation:
\begin{equation}
\gamma \frac{d^2 F}{d \xi^2} -(1-\gamma \Omega^2)F+ \left( 1+\Omega+iv \frac{d}{d \xi}
\right) |F|^2 F=0,
\end{equation}
where we have set $\gamma \equiv 1-v^2$. An inspection of the
asymptotic (linear) form of Eq.(8) as $\xi \rightarrow \infty$
shows that solitary waves exist provided that the condition
$(1-\gamma \Omega^2)/ \gamma >0$ is satisfied, which implies
$\gamma>0$, i.e. $|v|<1$ (slow gap solitons) and $| \Omega | < 1/
\sqrt{\gamma}$. For fixed values of the two family parameters
$\Omega$ and $v$, the homoclinic trajectory of Eq.(8) can be
numerically computed by standard techniques; the corresponding
magnetic field $\varphi$ can be then computed by numerical
integration of Eq.(5). Explicit analytical expressions of
companion electric and magnetic gap solitons have been found
solely in the steady case ($v=0$), which read:
\begin{eqnarray}
\psi(z',t') & = & [2(1-\Omega)]^{1/2} {\rm sech} [(1-\Omega^2)^{1/2} z'] \exp(i \Omega
t') \\
 \varphi(z',t') & = & -i \sqrt{\frac{\epsilon_{0}^{'}}{\mu_{0}^{'}}} \frac{\sqrt 2
(1-\Omega)}{\sqrt{1+\Omega}} \frac{\sinh[(1-\Omega^2)^{1/2} z']}{ \cosh^2
[(1-\Omega^2)^{1/2} z']} \exp(i\Omega t').
\end{eqnarray}
Note that the steady solitary waves (9) and (10) describe a
stationary {\it nonlinear localized mode} for the electromagnetic
field provided that its carrier frequency falls inside the bandgap
region ($|\Omega|<1$). Examples of electric and magnetic field
profiles, for a stationary [$v=0$; see Eqs.(9) and (10)] and
moving ($v \neq 0$) gap solitons, are shown in Fig.2. Note that,
in terms of real physical variables, spatial and temporal length
scales in the figures, for parameter values as in Fig.1, are
$5.77$ cm and $0.32$ ns, respectively. The soliton velocity $v$ is
measured in terms of the characteristic velocity
$v_g=(\omega_{0}^{2} \epsilon_0 \mu_0 \epsilon^{'}_{0}
\mu_{0}^{'})^{-1/2}$, whose value is $v_g \simeq 0.6 c_0$, where
$c_0=1/(\epsilon_0 \mu_0)^{1/2}$ is the speed of
light in vacuum.\\
An important issue is the stability of the solitary waves. A
rigorous stability analysis is challenging and can not be framed
in the analysis of Klein-Gordon solitons developed in
Ref.\cite{Laedke82_b}. However, from numerical integration of
Eqs.(5) and (6) we could ascertain stable propagation in a wide
range of the existence domain. In particular, we found that the
dispersive nonlinear term entering in Eq.(7) plays a {\it
stabilizing} effect on the solitary wave dynamics. We numerically
integrated Eqs.(5) and (6) using a pseudospectral split-step
technique with typical 512 discretization points in a traveling
reference frame $\xi=z'-vt'$ and $\eta=t'$, where the unperturbed
solitary wave is at rest; as an initial condition we assumed a
solitary wave perturbed with a small random noise to seed possible
instabilities. As an example, stable evolution of a steady
solitary wave, with carrier frequency at the band gap center, and
of a moving solitary wave are shown in Fig.3. We note that, if the
nonlinear dispersive term in Eq.(7) were neglected, the steady
solitary wave of Fig.3(a) would be unstable according to the
analysis of Ref.\cite{Laedke82_b}. We checked indeed the
stabilizing effect played by the nonlinear dispersive term by a
numerical analysis of Eq.(7) with and without the right hand side
term, assuming as an initial condition their respective non-moving
solitary waves with an added small random noise. We assessed the
reliability of our numerical method by exactly reproducing the
stability domain $1/ \sqrt 2 <|\Omega|<1$ for steady solitary
waves of the $\psi^4$ NLKGE according to the linear stability
analysis \cite{Laedke82_b}. Figure 4 shows the numerical results
of evolution of the steady solitary waves of Eq.(7) at the band
gap center in absence [Fig.4(a)] and in presence [Fig.4(b)] of the
nonlinear dispersive term. Note that, as in the former case the
solitary wave is unstable according to the linear stability
analysis \cite{Laedke82_b}, in the latter case no unstable growing
modes were observed in the numerical simulation, indicating that
the nonlinear dispersive term plays a stabilizing effect. The
emergence of instabilities for gap solitons and their physical
explanation are usually nontrivial issues and represent a
challenging task \cite{Barashenkov98}; therefore a detailed
physical explanation of the instability suppression observed in
our numerical simulations goes beyond the aim of the present work.
We just mention that for the NLKGE without the nonlinear
dispersive term the instability of the gap soliton close to the
band gap center ($|\Omega|< 1/ \sqrt 2$), as numerically
reproduced in Fig.4(a), arises because of the emergence of an
unstable internal mode of the soliton \cite{Laedke82_b}. Our
numerical results suggest that the presence of the nonlinear
dispersive term in the NLKGE makes such internal mode damped in
the entire bandgap region of linear waves ($|\Omega|<1$).

\section{Conclusions}
 In conclusion, we have predicted the existence of stationary and
moving gap solitary waves for electric and magnetic fields in
metamaterials with a Kerr nonlinearity. These waves, which are
supported by the nonlinear dielectric response of the medium,
exist in the band gap spectral region of the medium, corresponding
to $\epsilon<0$ and $\mu>0$, which separates the spectral regions
of allowed propagation where the medium behaves as a LH or as a RH
material. The gap solitary waves studied in this work - which can
be described by a complex NLKGE with a dispersive cubic term -
represent a new class of gap solitons in artificially-constructed
microstructured periodic media which do not involve Bragg
scattering and provide a noteworthy and physically relevant
example of self-transparent electromagnetic wave propagation in
the recently developed class of composite metamaterials.\\

\begin{figure}
\vspace*{20cm} \includegraphics{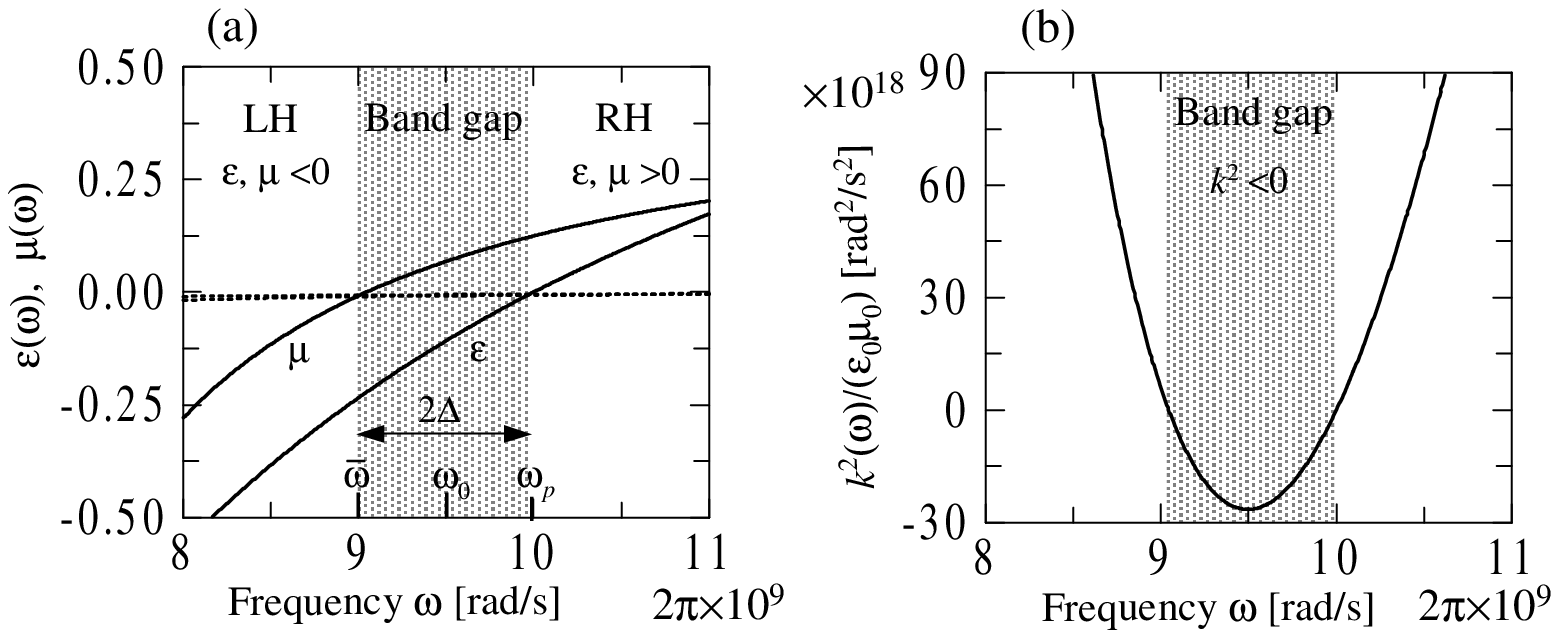}
\caption{Typical dispersion curves of a composite metamaterials with a stop band in the
 microwaves. (a) Behavior of relative dielectric permittivity $\epsilon(\omega)$ and
 magnetic permeability $\mu(\omega)$ versus frequency in the ideal lossless medium for
 parameter values $\omega_p=2 \pi \times 10$ GHz, $\omega_m= 2 \pi \times 6$ GHz and
 $F=0.56$ (corresponding to $\overline \omega \simeq  2 \pi \times 9$ GHz and a central bandgap
 frequency $\omega_0 \simeq 2 \pi \times 9.5$ GHz). The dashed
 curves (almost overlapped) in the figure show, for comparison, the behavior of imaginary
 parts of  $\epsilon$ and $\mu$ in a low-loss medium with damping terms
 $\gamma_{\epsilon}/\omega_p=\gamma_{\mu}/ \omega_p=0.005$. (b) Corresponding behavior of
 $k^2(\omega)$ versus frequency for linear waves in the lossless medium.}
\end{figure}

\begin{figure}
\vspace*{20cm} \includegraphics{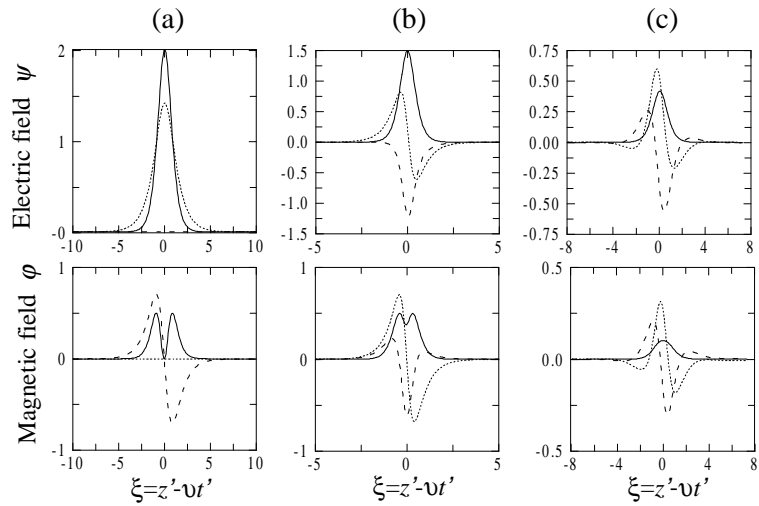}
\caption{Gap solitary waves for normalized electric field $\psi$ (upper figures) and
magnetic field $\varphi$ (lower figures) for a few values of family parameters $v$ and
$\Omega$. Continuous curves: modulus square of the fields ; dotted curves: real part of
the fields; dashed curves: imaginary part of the fields. In (a) steady soliton with
$v=0$ and $\Omega=0$ (see Eqs.(9) and (10) given in the text); in (b) moving soliton
with $v=0.8$ and $\Omega=0$; in (c) moving soliton with $v=0.8$ and $\Omega=1.2$ (the
real and imaginary parts are taken at $t'=0$ in this case). In the plots of lower
figures (magnetic field), we assumed $\epsilon^{'}_{0} / \mu^{'}_{0}=1$.}
\end{figure}

\begin{figure}
\vspace*{20cm} \includegraphics{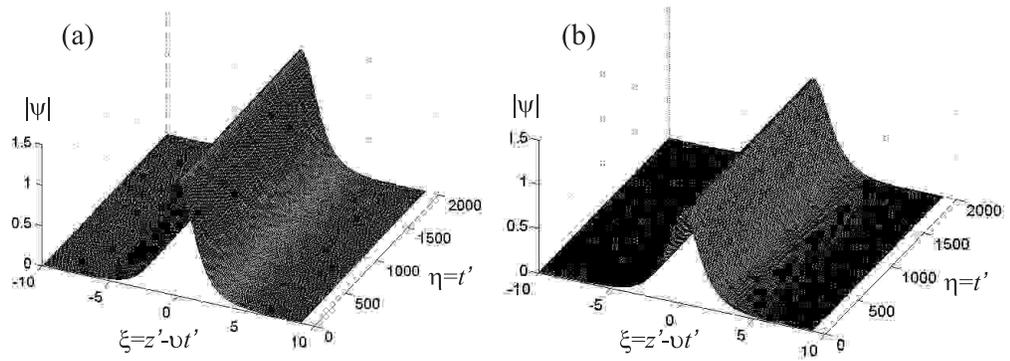}
\caption{Numerically computed evolution of gap solitons of Eqs.(5) and (6) in the
moving reference frame $\xi=z'-vt'$, $\eta=t'$ for (a) $v=\Omega=0$ and (b) $v=0.4$,
$\Omega=0.5$. Discretization points: 512. Time step: $d \eta=0.002$. Initial condition
is the solitary wave with an added small random noise (1$\%$ amplitude).}
\end{figure}

\begin{figure}
\vspace*{20cm} \includegraphics{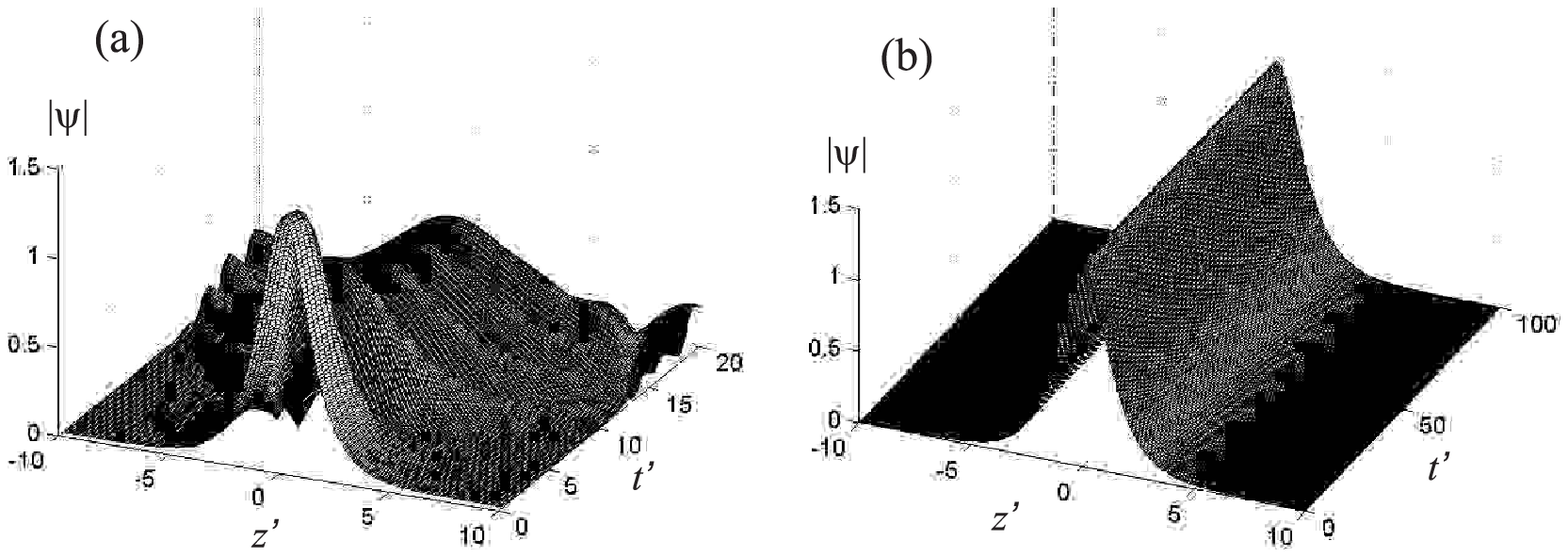}
\caption{(a) Temporal evolution of the stationary solitary wave of the $\psi^4$ complex
NLKGE [Eq.(7) with right hand side equal to zero] for $\Omega=0$ (time step
$dt'=0.001$, 512 discretization points). (b) Same as in (a), but for Eq.(7).}
\end{figure}


\begin{thebibliography}{99}
\bibitem{note0} It should be noted that, though artificially
constructed metamaterials show a periodic structure
\cite{Pendry99}, the interaction with the electromagnetic field
occurs in the long-wavelength limit (i.e. the wavelength of the
electromagnetic radiation is much larger than the characteristic
period of the microstructured material), so that a metamaterial
behaves as a homogeneous dielectric and magnetic medium with an
effective electric permittivity $\epsilon$ and magnetic
permeability $\mu$ which are strongly influenced by the internal
(microscopic) structure of the medium \cite{Pendry99}. The
existence of a frequency stop band arises because of $\epsilon
\mu<0$ in the frequency range separating LH and RH propagative
spectral regions. Conversely, in periodic media such as gratings
and photonic crystals the wavelength of the electromagnetic field
is comparable with the characteristic period of the structure and
Bragg diffraction effects are fundamental to create forbidden
spectral regions.
\bibitem{note2}
In fact, if the deviation of $\mu(\omega)$ from $\mu(\omega_0)$
were small for frequencies $\omega$ at around the stop band
spectral region, at leading order one could replace $(\omega_0-i
\partial_t) \mu(\omega_0-i\partial_t)$ with $\omega_0 \mu(\omega_0)$ in Eq.(3), and
in the corresponding NLKGE [Eq.(7)] the right-hand-side term would
disappear.
\bibitem{note1}
Homogeneous traveling waves of Eq.(7) turn out to be modulationally unstable, so that
dark or kink-type solitary waves are always unstable and hence not considered here.
\end{thebibliography}
\end{document}